\documentclass[preprint,showpacs,preprintnumbers,amsmath,amssymb]{revtex4}
\usepackage{graphicx}% Include figure files
\usepackage{dcolumn}% Align table columns on decimal point
\begin{document}
%\preprint{preprint - vortex group, smc/nims}
\title{Vortex matching effect in engineered thin films of NbN}

\author{Ajay D. Thakur$^{1}$, \footnote{Email : thakur.ajay@nims.go.jp}, Shuuichi Ooi$^1$, Subbaiah P. Chockalingam$^2$, John Jesudasan$^2$, Pratap Raychaudhuri$^2$, Kazuto Hirata$^1$}
\affiliation{$^1$ National Institute for Materials Science, 1-2-1 Sengen, Tsukuba, Ibaraki 305-0047, Japan\\
$^2$ Tata Institute of Fundamental Research, Homi Bhabha Road, Colaba, Mumbai 400005, India}
\date{\today}

\begin{abstract}
We report robust vortex matching effects in antidot arrays fabricated on thin films of NbN. The near absence of hysteresis between field sweep directions indicates a negligible residual pinning in the host thin films. Owing to the very small coherence length of NbN thin films ($\xi < 5~nm$), the observations suggest the possibility of probing physics of vortices at true nanometer length scales in suitably fabricated structures.
\end{abstract}

\maketitle

%\section{Introduction}

Vortex matching phenomena in superconducting wire networks and antidot arrays have been studied extensively in the past both as a model system for fundamental understanding of physics in the systems where interaction, disorder and thermal fluctuations mutually lead to interesting physical phenomena (e.g., multi-quanta vortex crystals, vortex ratchet effect, vortex n-mer states, etc. \cite{F1, M1, M2, C1, C2}) as well as from the point of view of possible applications (viz., possible route for enhancement of critical current densities, devices exploiting flux of vortices, i.e., fluxtronics, etc. \cite{C2, M3, M4, M5, M6, M7}). Most of these studies have concentrated on low transition temperature ($T_c$) superconductors Al, Nb and Pb \cite{F1, M1, M2, M3, M4, M5, M6, M7}. From the viewpoint of applications, whereas Al requires sub-Kelvin operational temperatures, both Nb and Al suffers from practical difficulties like oxidation in ambient atmospheric environments. Search for a suitable superconductor paving the way for a Pb free technology and working at temperatures accessible by closed cycle refrigerators is an important issue to realize viable applications based on the results of the exhaustive research on vortices in patterned superconductors. As thin films of NbN are chemically and structurally stable in ambient atmospheric environments and akin to their small coherence lengths ($\xi < 5~nm$), it is possible to fabricate structures at nanometer scales. They have been used in the past to fabricate bolometers for ultra sensitive calorimetric applications by detecting upto single photons to a high precision \cite{B1, B2, B3, B4, B5}. However, NbN has been considered as a dirty superconductor ($k_F~l~\sim~1$ within the Ioffe-Regel criteria \cite{I1}) and no reports exist in exploring the phenomena of vortex matching in superconducting wire networks and antidot arrays fabricated on NbN thin films in detail.

Epitaxial thin films of NbN grown on MgO substrates have recently been demonstrated \cite{P1, P2} as a suitable {\it s}-wave superconducting system for studying the interplay of carrier density and disorder on superconducting properties. It was shown that the carrier density plays a primary role in determining the $T_c$ of NbN thin films. In good agreement with Anderson theorem \cite{A1}, disorder scattering seems to play a negligible role in determining the $T_c$ of these films. The insensitivity of $T_c$ to disorder scattering suggests the possibility of a negligible bulk pinning in moderately clean samples having higher carrier densities with $k_F~l~\approx~7$. This motivated us to explore the vortex matching phenomena in engineered thin films of NbN containing patterned antidot arrays in various geometries.

The antidot array samples were prepared on NbN thin films grown on a (100) MgO substrate. For this, initially, a 60 nm thick NbN film was deposited through reactive dc sputtering using a Nb target in (1:5) Ar-N$_2$ partial pressure of 5 mTorr keeping the substrate at 600$^o$C. Further details of the film growth and characterization were published elsewhere \cite{P1}. We used a thin film with a $T_c^{on}$ (onset temperature for superconductivity transition) of 15 K and a $k_F~l$ of 5.9. A probe pattern is then made via photolithography using a mask-aligner followed by ion beam etching for a typical four probe measurement with region between the voltage probes typically being 40~$\mu$m x 40~$\mu$m wide. Within these 40~$\mu$m x 40~$\mu$m wide available regions, antidot arrays in suitable geometries were patterned using the Focussed Ion Beam (FIB) milling technique utilizing an aperture of 25~$\mu$m and a typical ion-beam current of 4.1 pA. Suitable mask patterns for the purpose were generated using the Micrion DMOD Description Language (mddl). The transport measurements were carried out via the conventional four-probe technique using a home made insert which goes into the cryostat of Quantum Design SQUID XL. The data acquisition was done utilizing the external device control (EDC) option. The temperature stability was better than 1~mK during the measurements.
%The etching was performed at a typical etching rate of 0.5 nm/sec in Argon plasma at an argon partial pressure of 0.75 mTorr with an antenna power of 100 W, a bias power of 100 W and the argon flow rate maintained at 2 cm$^3$/min. Under these etching conditions, there was essentially no damage to the NbN film. After removal of the photoresist,  

The NbN thin film used for the purpose of fabricating antidot arrays has a transition width of 100~mK, resistivity in normal state $\approx~1.48 \mu \Omega cm$, an electron mean free path of 3 $\AA$, a coherence length of 4.3 nm and a carrier density of $1.61 \times 10^{29}m^{-3}$. There is a reduction of $T_c$ upon the fabrication of antidot arrays by $\approx 100 mK$ and also an accompanying transition width broadening ($\Delta T_c \approx 0.63 K)$. Variation of resistance versus temperature measured using a dc drive current ($I_{dc}$) of 10 $\mu$A is shown for a typical antidot array in the main panel of Fig.1 (with $T_c$ marked). The inset in Fig.1 shows a scanning ion beam (SIB) image of the entire measurement geometry for the conventional four probe measurements. We next look at the phenomenon of matching effect in antidot arrays fabricated on a NbN thin film with different symmetries. Figure 2 shows the plot of {\it R} versus {\it f} in the case of a triangular antidot array at various reduced temperatures, $t$ ($=T/T_c$) close to $T_c$ (i.e., $t=1$) measured with $I_{dc}~=~100~\mu$A. Here, $f$ is the filling fraction, such that $f~=~H/H_1$, with $H_1 = 149.4~Oe$, where, $H$ is the applied magnetic field and $H_1$ is the first matching field. The corresponding SIB image of a portion of the triangular lattice is shown in the inset. Here, the distance between nearest antidots is $a_0$ ($=~400~nm$) and the average antidot diameter ($d$) is about 170~nm. From the main panel in Fig.2, it can be seen that a robust integer matching effect can be observed up to the eighth matching period at $t=0.931$. In Fig.3 we discuss the case of a square lattice of antidots with $a_0 = 350~nm$ and $d \approx 180~nm$. The corresponding SIB image of a section of the antidot lattice is shown in the inset of Fig.3. The main panel of Fig.3 shows a plots of {\it R} versus {\it f} at various $t$ values obtained with a dc drive current of $I_{dc}~=~100~\mu$A. The observed value of the first matching field turns out to be $H_1~\approx~169~Oe$. Even here one can unambiguously mark upto eighth matching period at $t=0.919$. 

Within the scenario proposed by Mkrtchyan and Shmidt \cite{MS1} we attempt to understand the above observations.
They showed that the maximum number of vortices that can be captured by an antidot (akin to a columnar defect) with a diameter $d$ is given by the saturation number ($n_s$) such that, $n_s = d/4 \xi (t)$, where $\xi (t)$ is the coherence length at reduced temperature $t$ given by $\xi (t) = \xi_0/ \sqrt{(1-t)}$, with $\xi_0$ being the zero temperature coherence length \cite{MS1}. In the case of the triangular antidot array with $d=180~nm$, $n_s$ lies in the range $2.60-2.36$ for $t$ lying in the range $0.931-0.943$. Similarly, in the case of square antidot array $n_s$ lies in the range $3.05-2.90$ for $t$ lying in the range $0.931-0.943$. These values of $n_s$ are far lower than the number of matching periods observed (see Figs. 1 and 2) and hence the multi-vortex scenario does not seem to account fully for the observations. It should however be noted that the saturation number obeys $n_s \sim (d/2 \xi(T))^2$ in the high field-regime because of the many external vortices which exert pressure into the antidots \cite{Doria}. Also the picture of Horng {\it et al} \cite{Horng} (where, there are missing matching fields due to the nucleation of interstitial vortices) does not seem to fit well to our data as we do not have any missing matching periods. Our observations could be better understood in the light of the scenario of ``super-matching" flux line lattices (SL's) \cite{Met1, Mart1, Vel1} formed by a reorganization of vortices that enters the interstitial sites at fields greater than $n_s H_1$. A characteristic feature of such a SL state is the existence of two kinds of fluxoids, viz., the vortices which are strongly trapped at the antidots and the vortices which are weakly pinned in the interstices of the antidot lattice. The depinning dynamics therefore depends on both the geometry of the antidot lattice as well as the direction of the current drive \cite{Met1} and is an interesting issue for future work.  

With a motivation of looking at the role played by intersitial vortices (the ones which sit at the interstitial sites between the antidots), we fabricated a honeycomb antidot lattice. The inset panel in Fig.4 shows the SIB image of a portion of the fabricated honeycomb lattice with the side of a honeycomb plaquette being $a_0$ ($=~400~nm$) and $d \approx 170~nm$. In the main panel of Fig.4, we show the results of the measurements on vortex matching phenomena in the case of the honeycomb lattice of antidots mentioned above. Plots of {\it R} versus {\it f} are shown at various drive currents in the range of $1~\mu$A to $150~\mu$A. As can be seen, we observe the conventional integer matching effect for filling fractions up to 3 with a matching period of 99.6~Oe. However, at $f=4$, we see an enhanced flux flow resistance. This can be understood within the qualitative scenario of interstitial vortices. Each antidot in the case of a honeycomb lattice is shared by three honeycomb plaquettes. Therefore a single vortex per antidot corresponds to a situation where there are two vortices per honeycomb plaquette. Now in our case, it appears that vortices go to the antidot sites up to $f=3$ (at $t=0.932$ we have $n_s \approx 2.57$). However, at $f=4$, the fourth vortex for each plaquette find it energetically more favourable to sit at the interstitial locations and hence at $f=4$, there are three vortices at each of the antidots whereas there are two vortices sitting at each of the interstitial positions. These interstitial vortices are more mobile and hence lead to enhanced flux flow resistance \cite{C4}.

In conclusion we have demonstrated robust vortex matching phenomena in antidot lattices of various geometries fabricated using FIB on NbN thin films. We also investigated the presence of vortex matching phenomena in relatively dirtier samples (with $T_c=10.9~K$ and $k_fl = 3.2$). However, preliminary results suggests an absence of robust signatures of the vortex matching phenomena pointing to the requirement for relatively clean samples for such observations. Further work is in progress to explore the role of sample disorder in more details. As NbN thin films have small coherence lengths ($\xi < 5~nm$) and are chemically and structurally stable in ambient atmospheric environments, they turn out to be an ideal system for studying vortex physics at true nanometer length scales as well as for future fluxtronics devices.

We have benefited from discussions with C. Reichhardt, T. Tamegai, A. K. Grover and S. Ramakrishnan. ADT and SO would like to acknowledge partial support from World Premier International Research Center (WPI) Initiative on Materials Nanoarchitectonics, MEXT, Japan. \\

%$^{\star}$~~thakur.ajay@nims.go.jp

\newpage

\begin{figure}

\caption{R versus T for a typical antidot array measured with a drive current of $10~\mu$A. The inset shows the SIB image of the entire measurement geometry.}

\caption{{\it R} versus {\it f} data obtained with a drive current $I_{dc}=100~\mu$A for a triangular antidot lattice observed in the temperature range of $0.931~T_c$ and $0.943~T_c$. The inset panel shows the SIB image of a portion of the triangular antidot lattice with a pitch of 400~nm and an average antidot diameter of 170~nm.}

\caption{{\it R} versus {\it f} data obtained with a drive current $I_{dc}=100~\mu~A$ for a square antidot lattice observed in the temperature range of $0.915~T_c$ and $0.923~T_c$. The inset panel shows the SIB image of a portion of the square antidot lattice with a pitch of 350~nm and an average antidot diameter of 180~nm.}

\caption{{\it R} versus {\it f} data obtained with a drive current in the range $I_{dc}=1~\mu$A~$-$~150~$\mu$A for a honeycomb antidot lattice observed at a temperature of $0.932~T_c$. The observed first matching period is $H_1=99.6~Oe$. The inset panel shows the SIB image of a portion of the honeycomb antidot lattice with a pitch of 400~nm and an average antidot diameter of 170~nm.}

\end{figure}

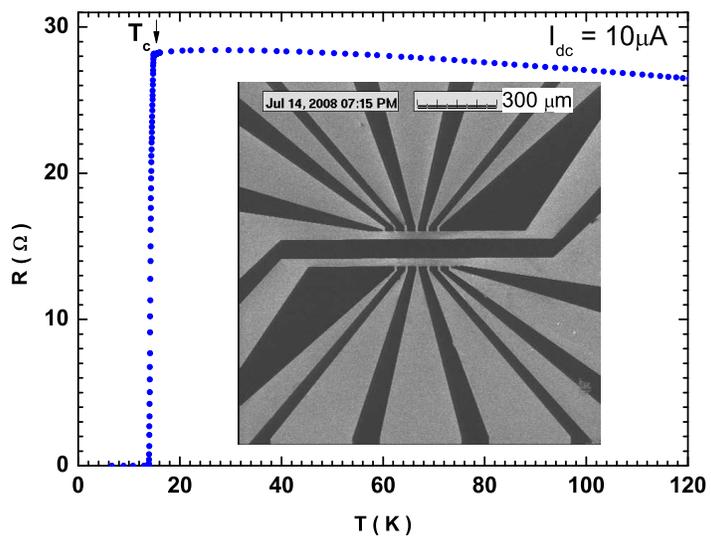
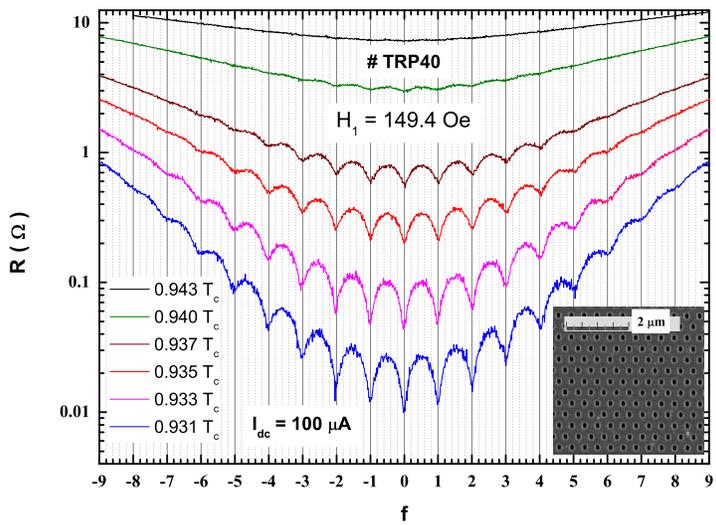
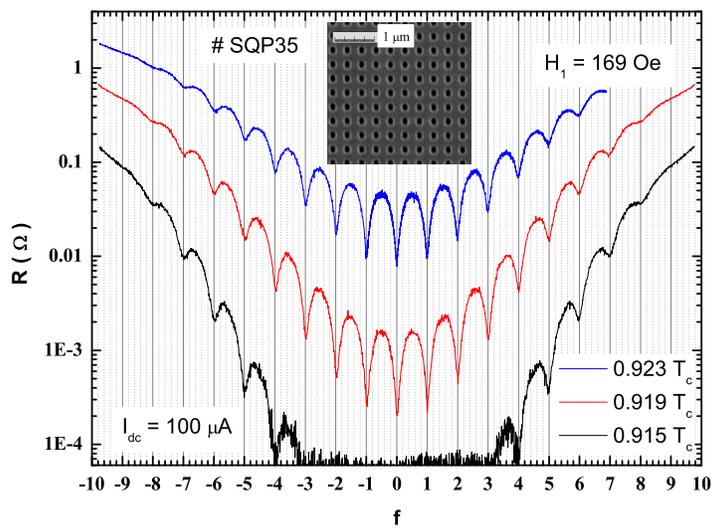
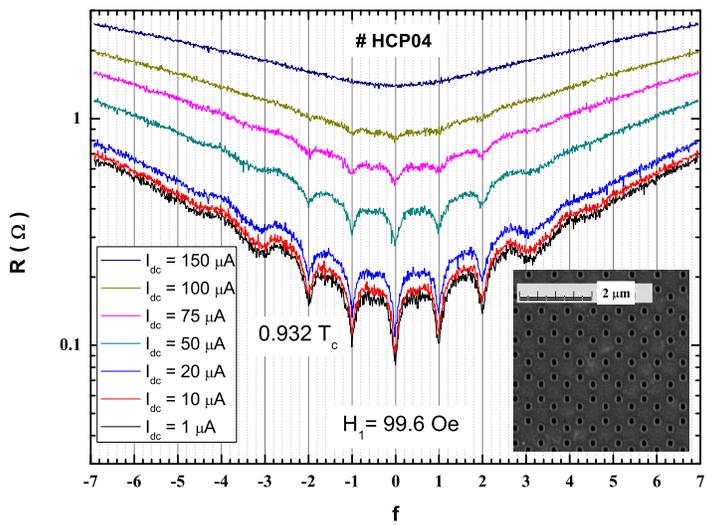
\end{document}